\documentclass[twocolumn,prl,showpacs,preprintnumbers,amsmath,amssymb]{revtex4}

\usepackage{amssymb}
\usepackage{graphics}
\usepackage{epsfig}
\usepackage{graphicx}    
\usepackage{dcolumn}     
\usepackage{amsthm}
\usepackage{bm}          
\usepackage{amsbsy}

\begin{document}
\title{Topological Transition of Graphene from Quantum Hall Metal
to Quantum Hall Insulator at $\nu=0$}
\author{W. Zhu$^{1,2}$, Q. W. Shi$^{1,2}$, J. G. Hou$^1$,
and X. R. Wang$^{2,3}$}
\email[Electronic address:]{phxwan@ust.hk}
\address{$^1$Hefei National Laboratory for Physical Sciences at
Microscale, University of Science and Technology of China, Hefei, P. R. China}
\address{$^2$Department of Physics, The Hong Kong University of
Science and Technology, Clear Water Bay, Kowloon, Hong Kong}
\address{$^3$School of Physics, Shangdong University,
Jinan, P. R. China}
\date{\today}

\begin{abstract}
The puzzle of recently observed insulating phase of graphene
at filling factor $\nu=0$ in high magnetic field quantum Hall
(QH) experiments is investigated. We show that the
magnetic field driven Peierls-type lattice distortion (due
to the Landau level degeneracy) and random bond fluctuations
compete with each other, resulting in a transition from a
QH-metal state at relative low field to a QH-insulator
state at high enough field at $\nu=0$. The critical field
that separates QH-metal from QH-insulator depends on the
bond fluctuation. The picture explains well why the field
required for observing the insulating phase is lower for
a cleaner sample.
\end{abstract}
\maketitle

\textit{Introduction. ---}The intriguing quantum Hall effect
(QHE) in graphene has attracted a lot of attentions in recent
years \cite{graphene,review}. The quantum Hall (QH) plateaus
was initially found to be \cite{QHE} $\sigma_{xy}=(4e^2/h)
(n+1/2)=\nu e^2/h, \ n=0,\pm 1,\pm 2,... $.
The result can be understood in the same manner as the QHE in
usual two-dimensional electron gases with graphene's special
properties \cite{QHE,Lee}: The factor of 4 comes from spin
and valley (so-called $K$ and $K'$ points) degeneracy, and
1/2, in the terminology of edge states, is from the
interesting splitting of $n=0$ Landau level (LL) at sample
edges due to the particle-hole symmetry \cite{Lee}.
This QHE rule was soon violated in stronger fields, 
and anomalous plateaus of $\nu= 0,\pm 1$
\cite{Jiang} due to the breaking of spin and valley
degeneracy were discovered although the detailed
symmetry-breaking mechanisms are still under debate \cite{Kun}.
The recent surprise comes from the discovery of insulating
phase at charge neutral point (CNP) in a high magnetic
field $B$ by Checkelsky \textit{et. al} \cite{Ong} because
it disagrees with the early appealing theory \cite{Lee}
that explained well all existing experiments then.
This discovery was confirmed by the temperature-dependence
study of magneto-transport near CNP (zero-energy) by Zhang
\textit{et. al} \cite{Yuan}. Experiments
\cite{Ong,Yuan} reveal following universal features.
1) Depending on the sample quality and applied field, graphene
at CNP in QH regime can be either a QH-metal or a QH-insulator (see below).
2) A magnetic field can drive a graphene at CNP from the
QH-metal state at relative low field into the QH-insulator
state at high enough field. The critical field required
for such a transition is lower for a higher quality sample.
3) The nature of the transition seem to be Kosterlitz-Thouless
type (KT-type).
\begin{figure}[b]
\includegraphics[width=0.45\textwidth]{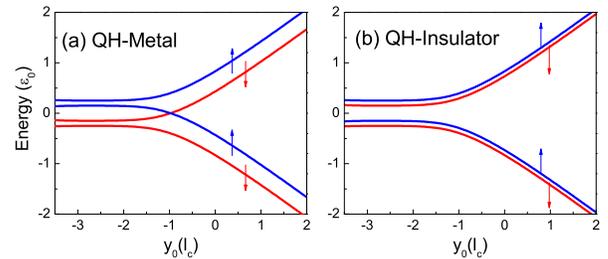}
\caption{(Color online) Schematic illustration of
position-dependent energy of $n=0$ LL for QH-Metal (a)
and QH-Insulator (b), respectively. The position (in unit
of magnetic length $l_c=\sqrt{\hbar/eB}$) is measured from
the sample edge ($y_0=0$) and $\epsilon_0$ represents energy
gap between $n=0$ and $n=1$ LLs. The blue and red lines
represent the spin-up and spin-down states, respectively.}
\end{figure}

Two possible $n=0$ states have been proposed and examined.
One is the \textit{QH-metal} \cite{Lee} that occurs when spin
split is larger than the valley split as illustrated in Fig.
1(a). The $n=0$ LL splits first into spin-up (blue curves) LL
and spin-down (red curves) LL. The further splits of valley
degeneracy near the sample edges create two counter-propagating
charge and spin carrying edge channels, which results in a
residual conductance of $2e^2/h$. This is a topological
state protected by the electron-hole symmetry \cite{Lee}.
The other possibility is the \textit{QH-insulator} when the
valley degeneracy lifts first with a smaller spin split
primarily due to the Zeeman effect. As shown in Fig. 1(b),
the $K-$LL and $K'-$LL move symmetrically to the opposite
direction of zero-energy (due to electron-hole symmetry).
As a result, the topological edge channels are absent at
zero energy, and a true insulator appears. The QH-insulator
was previously excluded from a possible $n=0$ state mainly
due to the early experimental observation of saturated
longitudinal resistance at low temperature \cite{Lee}.
However, the recent observation of insulating phase near CNP
suggests that QH-insulator is also a possible
zero-energy phase of graphene \cite{Ong,Yuan}.

The mechanism of the breaking of fourfold degeneracy of
$n=0$ LL has been discussed extensively in literature
\cite{Kun,Fisher,Gusynin,Igor,Koshino,distortion,Ryu}.
Whether graphene at zero-energy in the QH regime is described
by QH-metal or QH-insulator has also been studied by Yang and
Han \cite{Zhi}. They suggested that a graphene will be a
QH-metal if spin splits first while the graphene is a
QH-insulator if valley splits first.
However, the study did not address why the same
sample can change from a QH-metal to a QH-insulator.
Assume that there is a field-induced metal-to-insulator
transition, Nomura, Ryu, and Lee \cite{Nomura} tried to
identify possible order-parameter and to argue the transition
being the experimentally observed KT-type. In this letter,
we would like to provide a plausible path that a graphene at
CNP can change from a QH-metal state to a QH-insulator state
as the magnetic field increases. The hope is not only to
understand why both QH-metal and QH-insulator phases are possible
in the same sample, but also to provide a guide for experimental
manipulation of the transition between the two phases.
We will show that bond fluctuations due to intrinsic
ripples in a graphene can resist the $K$--$K'$ split in
low magnetic field, resulting in the QH-metal state at CNP.
In the opposite scenario, one should expect a QH-insulator state.

\textit{Picture.---} It shall be useful to first present our
picture of why a transition from the QH-metal to the QH-insulator
phases should be expected in graphene in a strong magnetic field.
Firstly, LLs of graphene in a strong magnetic field are highly
degenerated. According to Peierls instability for a solid or
Jahn-Teller effect for a molecule \cite{curtis}, the graphene can
lower its energy through a lattice distortion \cite{Lederer}.
The degree of distortion should be proportional to the magnetic
field because the LL degeneracy is proportional to the field
\cite{distortion,Lederer}. The distortion destroys $K$--$K'$
(or sublattices A and B for $n=0$ LL) degeneracy. According to
various estimation \cite{distortion,Nomura,Lederer,Hou},
the $K$--$K'$ split is linear in B-field for an ordered
distortion, similar to the usual Zeeman effect but bigger.
Although there may be other mechanism of breaking valley
degeneracy such as electron-electron interaction
\cite{Kun,Fisher,Gusynin}, but the Peierls instability
should be robust and universal for highly degenerated systems
like graphene that can also be regarded as a large molecule.
Secondly, it is inevitable to get ride of ripples in graphene.
The ripples create the intrinsic bond disorders, and the
intrinsic bond fluctuations tend to suppress the $K-K'$
split in $n=0$ LL because the bond fluctuation shall restore the
inversion symmetry of A- and B-sublattices \cite{Aoki,Guinea}.
As a result,  the lattice distortion and the intrinsic bond
fluctuations compete with each other. At relative low field,
the fluctuations overtake the distortion effect, and there is
no $K$--$K'$ valley splitting. The graphene at the Dirac point
is a QH-metal. However, at high enough field, the Peierls-type
lattice distortion dominates, and $K$--$K'$ splitting occurs.
The Dirac point becomes a QH-insulator.

\textit{Model and method. ---}Low energy excitations of
graphene are from $\pi-$electrons that can be modeled by
a tight-binding Hamiltonian on a honeycomb lattice,
$H=\sum\limits_{i\sigma}
\varepsilon_{i\sigma}|i\sigma><i\sigma|+\sum\limits_{<ij>,
\sigma}{(t_{ij}+\delta t_{ij})e^{i\phi_{ij}}
|i\sigma><j\sigma| +h.c.}$.
$t_{ij}$ is the hopping energy between two nearest-neighbor sites
whose value for a pure graphene is $t= -2.7eV$. $\sigma=\pm$ is
spin label. $\varepsilon_{i\sigma}$ is the on-site energy that
includes the Zeeman contribution $\pm g\mu_{B} B/2$ for spin-up ($+$)
and spin-down ($-$), leading to a spin-gap $\Delta_s=g\mu_B B=1.3
\times B[K/T]$.
One can also add a random on-site energy to mimic the extrinsic
disorders. $\delta t_{ij}$ is for the possible intrinsic random
hopping energy described later. The magnetic field is introduced
by Peierls' substitution in hopping parameter
$\phi_{ij}=2\pi e/h \int_i^j\vec{A}\cdot d\vec{l}$ \cite{xrw}.
The Peierls lattice distortion is either in-plane like Kekul\'{e}
bond order wave\cite{distortion,Nomura,Hou}, which alternates
the short and long bonds like in the classical benzene molecule
(see inset of Fig. 3(a)), or out-plane like charge density wave
(CDW) \cite{Lederer}, which breaks the inversion symmetry of
sublattices (see inset of Fig. 3(d)). The distortion is chosen
in such as way that valley gap is $\Delta_v=2.0\times B [K/T]$
\cite{distortion,Nomura,Hou} (for in-plane like Kekul\'{e}
bond order wave) or $\Delta_v=4.2\times B [K/T]$ \cite{Lederer}
(for out-plane CDW), both of which are bigger than $\Delta_s$.

\begin{figure}[b]
\includegraphics[width=0.45\textwidth]{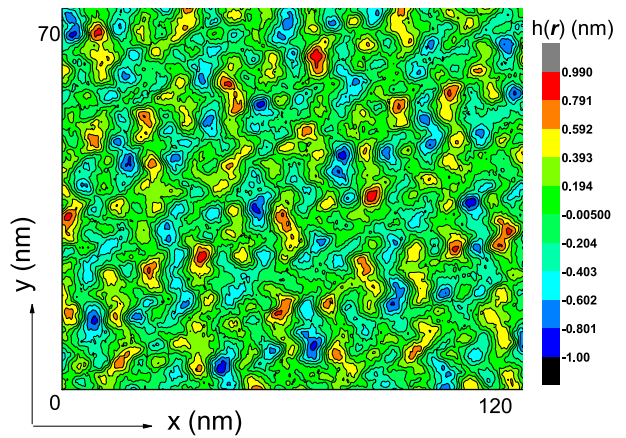}
\caption{(Color online) Contour plot of a randomly generated
landscape of the ripples used in our calculations. The average
lateral scale of ripple is about $\lambda_{ripple}=8nm$.
The height fluctuation is approximately in the range of
[-1.0,1.0]nm.}
\end{figure}

It is known that ripples (long-ranged corrugation) are the
intrinsic fluctuations of graphene \cite{Elena}. It is
observed that the graphene ripples have a height variation
of about $0.5\sim 1.0$nm, which occurs on a lateral scale of
$8\sim 10$nm. The graphene with random ripples
can be simulated by superposition of a series of plane
waves\cite{ripplesimulation}. The out-of-plane displacement is
$h(\textbf{r})=C\sum\limits_{i} C_{\textbf{q}_i}\sin
(\textbf{q}_i\cdot \textbf{r}+\delta_i)$,
where $\textbf{r}=(x,y)$ is the position of in-plane atoms
and $C_{\textbf{q}_i}$, $\textbf{q}_i$ and $\delta_i$ are
random numbers\cite{rippleparameter}.
The lowest possible $\textbf{q}_i$ determines lateral scale of
ripples, and the height fluctuation is controlled by $C$ value.
Fig. 2 is the contour plot of one corrugated graphene used in
our calculations. The random shapes and locations of ripples
lead to a random change in bond length that in turn causes
the hopping energy fluctuations \cite{review}: $\delta t_{ij}
=t\alpha \Delta a /a$, where $\Delta a=\sqrt{a^2+(h(\textbf
{r})-h(\textbf{r}'))^2}-a$ and $\alpha=\partial \log
t/\partial \log a =2$\cite{ripplesimulation}.

As it was explained in our early discussions,
whether a graphene is in QH-metal or QH-insulator states depends
on how the valley and spin split in the bulk. Thus one needs only
to study bulk states in order to determine two topologically
different states. To see how $n=0$ LL split into spin bands and valley bands in
the lattice distorted graphene with intrinsic random hopping, we
need to obtain an accurate density of states (DOS) of the model
numerically. The Lanczos recursive method \cite{Shang,zhu} on a
large lattice (with about million lattice sites) is employed.
The averaged DOS $\rho(E)= -\frac{1}{\pi}
Im<\psi|\frac{1}{E-H+i\eta}|\psi>$ is computed.
In the approach, a small artificial cut-off energy ($\eta=0.1meV$),
which results in a small LL width in clean graphene, is introduced
to simulate an infinitesimal imaginary energy \cite{zhu}.
A periodic boundary condition is used in order to minimize the
boundary effects since our concern is the bulk $n=0$ LL splits.

\textit{Results and discussions. ---}Fig. 3(a-c) are the DOS of
$n=0$ LL for in-plane distorted graphene of the Kekul\'{e} bond
order wave (shown in the inset of Fig. 3(a)) in the presence
of ripples shown in Fig. 2. At a relative low magnetic
field $B=9.4T$ (Fig. 3(a)), the distortion-induced valley
gap cannot compete with the hopping fluctuation so that bond
order wave is statistically destroyed. As a result, $K$--$K'$
degeneracy is statistically restored, and $n=0$ LL splits into
spin-up (blue) and spin-down (red) bands due to the Zeeman
energy as shown in Fig. 3(a). This is the QH-metal state
discussed early because the spin splitting dominates the bulk
gap of $n=0$ LL. As the magnetic field is above the critical
field $B_c$ (between $9.4$T and $13.2$T in our case as indicated
in Fig. 3(a-b)), the distortion-induced valley gap appears and
$n=0$ LL splits into four bands.
As shown in Fig. 3(b), two peaks on the left (right) sides of
zero energy are for $K$ ($K'$) valleys and the corresponding edge
states do no cross each other. The graphene is in the QH-insulator
state at CNP. Further increase of the field, the valley splits
become even bigger as shown in Fig. 3(c) for $21.9$T.
To shown that above result is very robust against different type of
distortion, we made the same type of calculations for the CDW lattice
distortion  (shown in the inset of Fig. 3(d)) \cite{Lederer}.
The similar results as those for the  Kekul\'{e} bond wave have been
found. As shown in Fig. 3(d-f). The valley degeneracy is preserved
at $6.6$T (d) and the zero-energy state is a QH-metal. The valley
degeneracy is broken at $11.0$T (e), resulting in a QH-insulator at CNP.
Further increase of field to $21.9$T (f) in this case, the two $K$ ($K'$)
bands are clearly located in the same side of the zero-energy point.
\begin{figure}[t]
\includegraphics[width=0.45\textwidth]{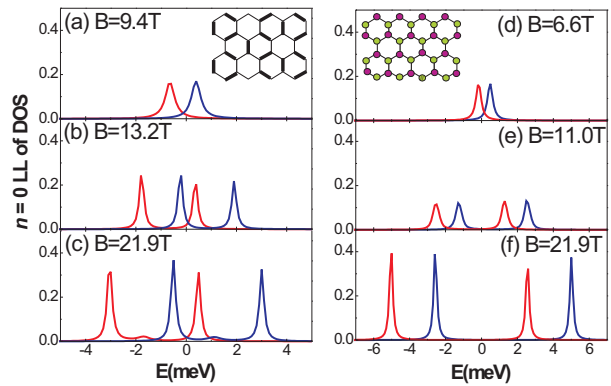}
\caption{(Color online) Density of state of $n=0$ LL of lattice
distorted graphene in the presence of intrinsic random hopping
for various magnetic field. (a-c) are for in-plane distortion of
Kekul\'{e} bond wave in $B=$9.4T (a); 13.2T (b); and 21.9T (c).
(d-f) are for out-plane CDW with $B=$6.6T (d); 11T (e); and 21.9T (f).
The blue (red) line represents the spin-up (spin-down) state.}
\end{figure}

Fig. 3 shows clearly that the intrinsic bond randomness due to
ripples can suppress lattice distortion-induced valley splitting no
matter whether the distortion is in-plane (Kekul\'{e} bond wave)
or out-plane (CDW), demonstrating the robustness and universality of
the picture for grapheme. As shown from our calculations, the valley
degeneracy is destroyed and valley gap appear when the bond disorder
is not strong enough to restore inversion symmetry of $A$--$B$
sublattices statistically. A cleaner sample should have a weak bond
disorder, and the corresponding valley degeneracy can be broken by
a smaller lattice distortion. In another word, the critical
field $B_c$ that induces the transition from the QH-metal state to
the QH-insulator state depends on the degree of bond randomness.
The cleaner a sample is, the smaller the critical field required to
drive the graphene from the QH-metal into the QH-insulator is needed.
In our model the bond randomness is measured by the height-fluctuation
$h_{max}$ of ripples. Fig. 4 is $h_{max}$-dependence of the critical
field $B_c$ for both Kekul\'{e} type and CDW-type lattice distortion.
It is found that the value of $B_c$ of CDW type is lower than that
of Kekul\'{e} type because the valley gap in CDW distortion is bigger.
The inset of Fig. 4 is the phase diagram for the QH-insulator and the
QH-metal. This is exactly what was observed in experiments \cite{Ong,Yuan}.
This also explains that failure of the discovery of the zero-energy
insulating phase at high field in experiments before Checkelsky
\textit{et. al} \cite{Ong} is probably due to the quality of the samples.

\begin{figure}[t]
\includegraphics[width=0.45\textwidth]{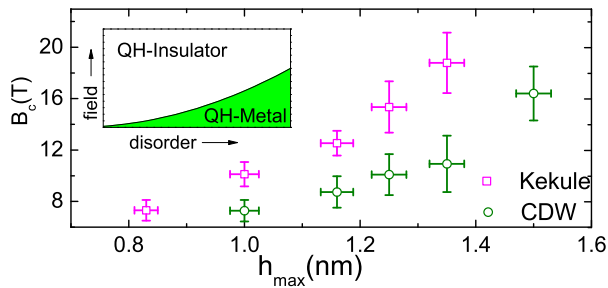}
\caption{(Color online) Relationship between critical field $B_c$
and ripple effects for Kekul\'{e} bond order (magenta) and CDW (olive),
respectively. $h_{max}$ represents the maximum height fluctuation
of spatial ripples of samples. Inset: Phase diagram of the
QH-metal state and QH-insulator state.}
\end{figure}

We would like to make a few remarks before ending this paper.
1) Fixed the applied magnetic field, the random fluctuation drives
a graphene from a zero-energy insulator to a zero-energy metal.
This is opposite to the role of randomness usually played in a
metal-to-insulator transition.
2) There are two types of disorders in graphene, the intrinsic bond
disorder due to the spontaneous ripple formation and extrinsic
disorder from the impurity states in substrate
\cite{Tan} and/or ad-atoms \cite{review}.
The extrinsic disorder can cause the on-site-energy randomness
in the tight-binding model. Our numerical studies show that the
on-site disorders will not suppress the $K$--$K'$ split,
but merely broaden $K-$ and $K'-$ subbands \cite{Koshino,onsite}.
The Landau subband broadening can cause inter-Landau-subband mixing that
may change the localization property \cite{Xiong} of a disordered system.
3) Our QH-metal is in fact a topological insulator because
it is a bulk insulator with conducting edge channels\cite{Lee} while
the QH-insulator is a conventional insulator.
The present work is on why the magnetic field can drive such a
topological transition rather than the nature of the transition
\cite{Nomura}.
4) The Coulomb interaction is neglected here because
we do not expect the change of physics by the interaction.

We present a theory of graphene for the field-driven topological
transition from the QH-metal state to the QH-insulator state at
$\nu=0$. The transition results from the competition between
Peierls-type lattice distortion and random bond fluctuation.
Our theory provides not only a clear explanation about the
existence of both QH-metal state at low field and QH-insulator
state at high field near the CNP, but also why the critical field
for a cleaner sample is lower.

This work is supported by Hong Kong RGC grants (\# 604109,
HKU10/CRF/08- HKUST17/CRF/08 and SBI07/08.SC09); the NNSF of China
(No. 10974187); NKBRP of China (No. 2006CB922005); and KIP of the
Chinese Academy of Sciences (No. KJCX2-YW-W22).


\begin{thebibliography}{99}
\bibitem{graphene}K. S. Novoselov, \textit{et. al}, Science {\bf 306}, 666(2004).
\bibitem{review}A. K. Geim, \textit{et. al}, Nat. Mater. {\bf 6}, 183(2007);
A. H. Castro Neto, \textit{et. al}, Rev. Mod. Phys. {\bf 81}, 109(2009);
S. Das Sarma, \textit{et. al}, arxiv-1003.4731.
\bibitem{QHE}K. S. Novoselov, \textit{et. al}, Nature {\bf 438}, 197(2005);
Y. Zhang, \textit{et. al}, Nature, {\bf 438}, 201(2005).
\bibitem{Lee}D. A. Abanin, \textit{et. al}, Phys. Rev. Lett. {\bf 96}, 176803(2006);
D. A. Abanin, \textit{et. al}, {\it ibid.} {\bf 98}, 196806(2007).
\bibitem{Jiang}Y. Zhang, \textit{et. al}, Phys. Rev. Lett. {\bf 96}, 136806(2006);
Z. Jiang, \textit{et. al}, {\it ibid.} {\bf 99}, 106802(2007).
\bibitem{Kun}K. Yang, Solid State Communications {\bf 143}, 27(2007);
S. Das Sarma and K. Yang, {\it ibid.} {\bf 149}, 1502(2009).
\bibitem{Ong}J. G. Checkelsky, \textit{et. al}, Phys. Rev. Lett. {\bf 100},
206801(2008);J. G. Checkelsky, \textit{et. al}, Phys. Rev. B {\bf 79}, 115434(2009).
\bibitem{Yuan}Liyuan Zhang,  \textit{et. al}, arxiv-1003.2738.
\bibitem{Fisher}Jason Alicea, \textit{et. al}, Phys. Rev. B {\bf 74}, 075422(2006).
\bibitem{Gusynin}V. P. Gusynin, \textit{et. al}, Phys. Rev. B {\bf 74}, 195429(2006);
E. V. Gorbar, \textit{et. al}, {\it ibid.} {\bf 66}, 045108(2002).
\bibitem{Igor}Igor A. Luk'yanchuk, \textit{et. al}, Phys.
Rev. Lett. {\bf 100}, 176404(2008).
\bibitem{Koshino}M. Koshino and T. Ando, Phys. Rev. B {\bf 75}, 033412(2007);
N. H. Shon, \textit{et. al} J. Phys. Soc. Jpn. {\bf 67}, 2421(1998).
\bibitem{distortion}N. A. Viet, \textit{et. al}, J. Phys. Soc. Jpn. {\bf 63},
3036(1994);H. Ajiki and T. Ando, {\it ibid.} {\bf 64}, 260(1995).
\bibitem{Ryu}Shinsei Ryu, \textit{et. al}, Phys. Rev. B {\bf 80}, 205319(2009).
\bibitem{Zhi}Z. Yang, \textit{et. al}, Phys. Rev. B {\bf 81}, 115404(2010).
\bibitem{Nomura}K. Nomura, \textit{et. al}, Phys. Rev.
Lett. {\bf 103}, 216801(2009).
\bibitem{curtis}L. Curtiss, \textit{et. al}, Phys. Rev. Lett.
{\bf 69}, 2435(1992).
\bibitem{Lederer}Jean No\"{e}l Fuchs, \textit{et. al},
Phys. Rev. Lett. {\bf 98}, 016803(2007);
J. N. Fuchs, \textit{et. al}, Eur. Phys. J. Special Topics {\bf 148}, 151(2007).
\bibitem{Hou}C. Y. Hou, \textit{et. al}, Phys. Rev. B {\bf 81}, 075427(2010).
\bibitem{Aoki}Tohru Kawarabayashi, \textit{et. al}, Phys. Rev. Lett. {\bf 103}, 156804(2009);
Tohru Kawarabayashi, \textit{et. al}, Physica E, {\bf 42}, 759(2010).
\bibitem{Guinea}F. Guinea, \textit{et. al}, Phys. Rev. B {\bf 77}, 205421(2008).
\bibitem{xrw}X. R. Wang, Phys. Rev. B {\bf 51}, 9310(1995); {\it ibid.} {\bf 53}, 12035(1996);
R. Saito, G. Dresselhaus and M. S. Dresselhaus,
\textit{Physical Properties of Carbon Nanotubes}, Imperial College Press,
London(1998).
\bibitem{Elena}Elena Stolyarova, \textit{et. al}, Proc. Nat. Aca. Sci. {\bf 104}, 9209(2007);
Jannik C. Meyer, \textit{et. al}, Nature, {\bf 446}, 60(2007);A. Fasolino, \textit{et. al}, Nat. Mater. {\bf 6}, 858(2007).
\bibitem{ripplesimulation}J. W. Klos, \textit{et. al}, Phys. Rev. B {\bf 80}, 245432(2009);
S. Costamagna, \textit{et. al}, Phys. Rev. B {\bf 81}, 115421(2010).
\bibitem{rippleparameter}The wave vectors $\textbf{q}_i$ randomly distributed in the range
$[2\pi/L,2\pi/3a]$, where $L$ is the typical length of the system and $a=1.42{\AA}$ is the
lattice constant. The phase $\delta_i$ is randomly distributed in the range
$[0,2\pi]$. $C_{q_i}=\sqrt{2}/q^2_i$ for $q_i>q_{ripple}$ and otherwise $C_{q_i}=\sqrt{2}/q^2_{ripple}$,
where $q_{ripple}=2\pi/\lambda_{ripple}$, $\lambda_{ripple}\approx 8nm$ represents the
averaged distance of the adjacent peaks of spatial ripples. In our calculations,
there are $>100$ random waves to construct the spatial ripples.
\bibitem{Shang}Shangduan Wu, \textit{et al.}, Phys. Rev. B {\bf 77},
195411(2008); and the references there in.
\bibitem{zhu}W. Zhu, \textit{et. al}, Phys. Rev. Lett. {\bf 102}, 056803(2009).
\bibitem{Tan}Y. -W. Tan, \textit{et. al}, Phys. Rev. Lett. {\bf 99}, 246803(2007).
\bibitem{onsite}We also calculate effects of on-site disorder on the DOS
of $n=0$ LL for in-plane Kekul\'{e} bond order and out-plane CDW order,
respectively. Our results show that on-site disorder can broaden both
spin- and valley-resolved $n=0$ Landau subband \cite{zhu}, but it cannot prevent
valley split.
\bibitem{Xiong}G. Xiong, \textit{et. al}, Phys. Rev. Lett. {\bf 87}, 216802(2001).
\end{thebibliography}
\end{document}